\documentclass[fleqn,usenatbib]{mnras}

\usepackage{newtxtext,newtxmath}

\usepackage[T1]{fontenc}
\usepackage{ae,aecompl}

\usepackage{graphicx}	
\usepackage{amsmath}	
\usepackage{amssymb}	

\begin{document}


\title[Intensity interferometry with MAGIC]{Optical intensity interferometry observations using the MAGIC imaging atmospheric Cherenkov telescopes}

\author[V.~A.~Acciari~et.~al.]{
V.~A.~Acciari$^{1}$, M.~I. Bernardos$^{2}$, E. Colombo$^{1}$, 
J. L. Contreras$^{3}$
J. Cortina$^{2}$\thanks{E-mail: juan.cortina@ciemat.es},
\newauthor
C. Delgado$^{2}$, 
C. D\'{\i}az$^{2}$, 
D. Fink$^{4}$, 
M. Mariotti$^{5}$,
S. Mangano$^{2}$\thanks{E-mail: salvatore.mangano@ciemat.es}, 
R. Mirzoyan$^{4}$,
\newauthor
M. Polo$^{2}$, 
T. Schweizer$^{4}$\thanks{E-mail: tschweiz@mppmu.mpg.de}
M. Will$^{4}$.
\\
$^{1}$ {Inst. de Astrof\'isica de Canarias, E-38200 La Laguna \& Universidad de La Laguna, Dpto. Astrof\'isica, E-38206 La Laguna, Spain} \\
$^{2}$ {Centro de Investigaciones Energ\'eticas, Medioambientales y Tecnol\'ogicas, E-28040 Madrid, Spain}\\
$^{3}$ {IPARCOS Institute and EMFTEL Department, Universidad Complutense de Madrid E-28040 Madrid, Spain}\\
$^{4}$ {Max-Planck-Institut f\"ur Physik, D-80805, Munich, Germany}\\
$^{5}$ {Universit\`a di Padova and INFN, I-35131 Padova, Italy} 
}
\date{Accepted XXX. Received YYY; in original form ZZZ}

\pubyear{2019}

\label{firstpage}
\pagerange{\pageref{firstpage}--\pageref{lastpage}}
\maketitle

\begin{abstract}
Imaging Atmospheric Cherenkov Telescopes (IACTs) currently in operation feature large mirrors and order of 1 ns time response to signals of a few photo-electrons produced by optical photons. This means that they are ideally suited for optical interferometry observations. Thanks to their sensitivity to visible wavelengths and long baselines optical intensity interferometry with IACTs allows reaching angular resolutions of tens to microarcsec. We have installed a simple optical setup on top of the cameras of the two 17 m diameter MAGIC IACTs and observed coherent fluctuations in the photon intensity measured at the two telescopes for three different stars. The sensitivity is roughly 10 times better than that achieved in the 1970's with the Narrabri interferometer.
\end{abstract}




\section{Historical introduction}

Hanbury-Brown \& Twiss\citep{HBT1954, HBT1956}  demonstrated that at sufficiently short baselines, the spatial correlation measurements of thermal light sources, such as stars, should exhibit a photon bunching signal for measurements within a defined area on the ground. This area depends on the angular size of the star so sampling the intensity of the correlated signal allows measuring the size of the star.

After initial tests on Sirius a two 6.5 meter diameter telescope interferometer with adjustable baseline was built in Narrabri, Australia \citep{HB1974}. This interferometer was used to measure the diameters of the 32 brightest stars in the Southern Hemisphere. Because the interferometer operated at visible wavelengths and the distance between the two telescopes was as large as 188 m Narrabri's interferometer could measure diameters as small as $\sim$400 $\mu$arcseconds.

Such ''intensity interferometry'' has hardly developed in the following decades probably because, contrary to ''amplitude interferometry'' \citep{gravity, chara}, it calls for large photon collection areas, sensitivity to a few photons, and fast time response in the order of a few ns down to ps. The situation may have changed recently due to the availability of telescopes with similar requirements: the so-called Imaging Atmospheric Cherenkov telescopes (IACTs). IACTs \citep{Hillas} detect faint (few photons) and fast (ns) pulses of near UV to blue Cherenkov light emitted by particle showers produced by cosmic or gamma rays above tens of GeV. Current and future IACTs may be modified to work as intensity inferferometers \citep{Dravins, VERITAS, CTA}.

Besides, conventional photomultipliers (PMTs) have time resolutions exceeding 1~ns but other photodetectors are even faster, potentially reaching $<$1~ns resolutions. Avalanche Photo Diodes have been recently applied to intensity interferometry \citep{Guerin}. 

\section{Short theoretical review}

The reader can find a detailed discussion on the theory of the intensity interferometer and the methods to analyze the correlation signal in the book by Hanbury-Brown \citep{HB1974}. Here we will only review a few definitions and expressions that are essential for our analysis.

Consider two telescopes pointing at a star, as sketched in figure \ref{fig:interferometer_sketch}. Let I$_1$(t) and I$_2$(t) be the signal intensities registered at each of the telescopes as a function of time. 

The second order coherence function g$^{(2)}$ is defined as:
\begin{equation}
    g^{(2)} = \frac{<I_1(t) \cdot I_2(t+\tau)>}{<I_1(t)> \cdot <I_2(t)>}
	\label{eq:g2_definition}
\end{equation}
where $<>$ represents averaging over time and $\tau$ is a time delay introduced by the different light and signal paths in the two telescopes.

For a randomly polarized light source and a light detector whose electronic bandwidth $\Delta f$ is much smaller than the optical bandwidth $\Delta\nu$ it can be shown that:
\begin{equation}
    g^{(2)} = 1 + \frac{\Delta f}{\Delta \nu} |V_{12}|^2
	\label{eq:g2_and_V}
\end{equation}
where the complex visibility V$_{12}$ is the Fourier transform of the source brightness distribution. 

In the remaining of this work we will rather use the normalized contrast $c$, defined as:
\begin{equation}
    c = \frac{ < (I_1(t)-<I_1>) \cdot (I_2(t+\tau)-<I_2>) >}{<I_1(t)> \cdot <I_2(t)>}
	\label{eq:c_definition}
\end{equation}

It is straightforward to show that:
\begin{equation}
    c = g^{(2)} - 1 = \frac{\Delta f}{\Delta \nu} |V_{12}|^2
	\label{eq:c_and_g2}
\end{equation}

For a star that is described as a disk of uniform intensity up to an angular diameter $\theta$ it can be proven that:
\begin{equation}
    |V_{12}| = 2 \frac{B_1(\pi \cdot d \cdot \theta / \lambda)}{\pi \cdot d \cdot \theta / \lambda}
	\label{eq:bessel}
\end{equation}
where B$_1$ is the Bessel function of the first order, $\lambda$ is the light wavelength and $d$ is the distance between the telescopes (baseline).

Dividing $c$ for a given baseline $d$ and $c$ at baseline 0 (or a small enough baseline) returns |V$_{12}$(d)|$^2$:
\begin{equation}
    \frac{c(d)}{c(0)} = |V_{12}(d)|^2
	\label{eq:cd_c0}
\end{equation}

$c$(0) is in effect a constant of the setup.

In conclusion $c$ decreases with increasing baseline following Eq. \ref{eq:bessel}. A measurement of the curve $c(d)/c(0)$ vs $d$ allows us to determine the angular size of the star.

\section{Description of MAGIC and the technical modifications to implement the interferometer}

MAGIC is a system of two IACTs located at the Roque de los Muchachos Observatory (Observatorio del Roque de los Muchachos, ORM) on the island of La Palma in Spain \citep{upgrade1}. Equipped with 17~m diameter mirror dishes and fast PMT cameras, the telescopes record images of extensive air showers in stereoscopic mode, enabling the observation of VHE $\gamma$-ray sources at energies $\gtrsim 50$ GeV \citep{upgrade2}. 

The telescope reflector follows a parabolic shape to minimize the time spread at the focal plane. The focal ratio is 1. The reflector is formed by $\sim$250 1~m$^2$ spherical mirror tiles. Two actuators behind each tile can be used to shift its orientation and correct the reflector for deformations in its parabolic shape as the telescope changes zenith. 
This so-called Active Mirror Control (AMC) adjusts the mirror in a few seconds and runs typically every 20 minutes during standard observations. Approximately 70\% of the light of a point source is focused on a pixel. It must be noted that individual mirror tiles are slightly staggered thus introducing an extra time spread in the order of 200~ps.

Each telescope is equipped with a 1039-pixel PMT camera at the primary focus. The PMTs are 25.4~mm diameter. A hexagonal shape Winston cone is mounted on top on each PMT. The distance between PMT centers is 30~mm. Each pixel covers a 0.1$^{\circ}$ FOV. 

The PMTs are Hamamatsu, type R10408, with a hemispherical photocathode and 6 dynodes. The PMT bias voltages for the cathode and dynodes are generated by a Cockroft-Walton DC-DC converter, which can provide up to 1250~V peak voltage. The electrical signals are amplified (AC coupled, $\sim$25 dB amplification) and then transmitted via independent optical fibers by means of vertical cavity surface emitting lasers (VCSELs). The average pulse width signal is measured to be 2.5~ns (FWHM). 

The PMTs operate at a rather low gain of typically 3-4 $\times$ 10$^4$ in order to also allow observations under moonlight without damaging the dynodes. 

\begin{figure}
	\includegraphics[width=\columnwidth]{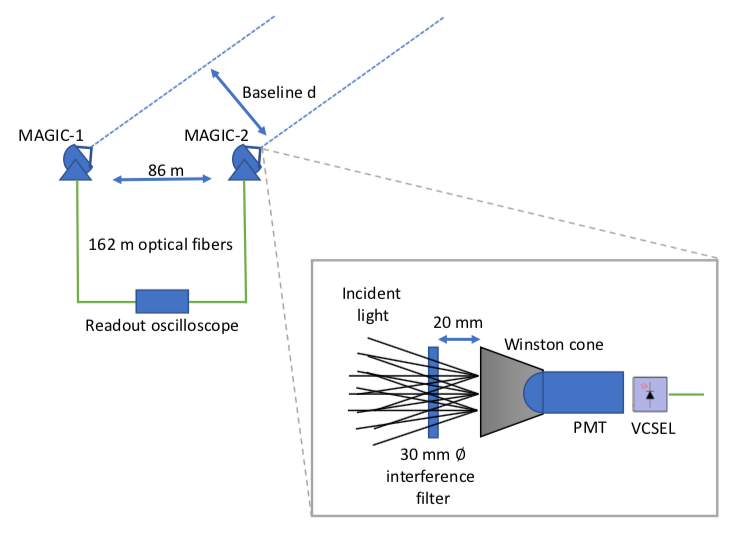}
    \caption{Sketch of the MAGIC telescopes and the new components that have been installed 
    to enable it as an optical intensity interferometer.}
    \label{fig:interferometer_sketch}
\end{figure}

Fig. \ref{fig:interferometer_sketch} shows a sketch of the two MAGIC telescopes along with a few of the elements that were installed for the interferometric observations.

For interferometry observations we only use the signal from the central PMT of each of the two telescopes. We operate the detector in so-called ''analog mode'', that is, we continuously sample and record the signal at the PMTs in a regime where each $\sim$2-4 ns sample corresponds to tens of photo-electrons. The sensitivity of the detector is proportional to the photon flux detected by the PMT and in first approximation does not depend on the optical bandpass of the detected light (see equation \ref{eq:significance} below).

However we have added a filter in front of the central pixels to make the anode currents of the PMTs (DCs) low enough that PMTs are not damaged when pointing at bright stars or during bright Moon observations. The filter is held 15~mm in front of the Winston cone using a mechanical frame previously developed for bright Moon VHE observations \citep{moon}. This frame must be mounted and dismounted manually, which generally prevents standard VHE observations for the whole night. 

\begin{figure}
	\includegraphics[width=\columnwidth]{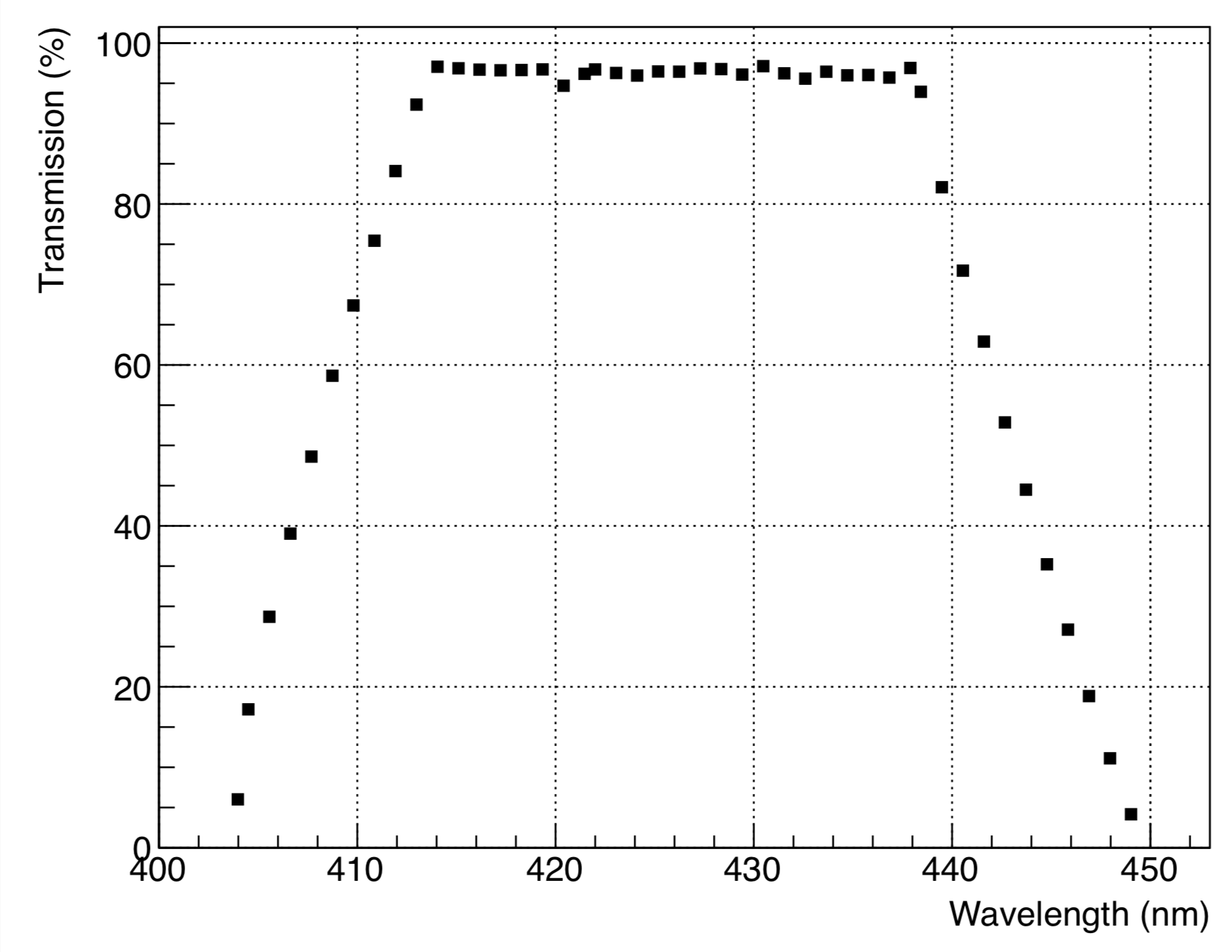}
    \caption{Effective transmission of the filter installed in front of the MAGIC central pixels as a function of wavelength, as calculated using a Monte Carlo simulation of the optical system.}
    \label{fig:effective_transmission}
\end{figure}
We selected the interference filter Semrock 432/36 nm BrightLine of 32~mm diameter and 3.5 mm thickness. The filter transmission is $>$90\% for 414-450~nm. However light enters the filter directly from the mirror so it follows a wide range of incident angles up to 26.6$^{\circ}$. As a result the effective filter transmission curves shifts to shorter wavelength. The filter has a mild dependence on incident angle (effective refractive index 2.02 for both polarizations). Fig. \ref{fig:effective_transmission}
shows the effective transmission range as obtained with a simulation that takes into account the shape of the telescope reflector, the geometry of both filter and pixel, the filter transmission for normal incidence and the dependence on transmission wavelength with incident angle. The center wavelength shifts to $\sim$427 nm and the FWHM of the transmission curve is roughly the same as the original one but the transmission curve develops two tails at short and long wavelengths. This actually reduces the sensitivity of the interferometer by about 15\% (see section \ref{sec:selection}). 

During normal VHE observations the reflector is focused at a distance of $\sim$10 km so that the images of atmospheric showers produced by $\gamma$-rays are as sharp as possible. For our observations we would rather focus the telescope at the star i.e. at infinity. AMC allows to move the focus in all three directions. We shifted it along the optical axis in small steps of a few mm and measured DC in the central pixels until we reached a maximum DC when focusing about 33~mm beyond the nominal focus for VHE observations. The filter has essentially the same physical size of a Winston cone and is situated some 20~mm ahead of it so a fraction of the light is shadowed. A simulation shows that this fraction is 18\%. This fraction was confirmed comparing DC measurements with and without the filter frame.

The MAGIC cameras transmit the analog signal of each pixel as an optical signal via fiber optics to a separate readout location. The fibres are $\sim$162~m long.
The analog optical signal from each of the two telescopes is generated by means of a multimode VCSEL present in each pixel of the camera. The 780~nm multimode optical signal is converted to an electrical signal at the readout end of the fiber using a biased photodiode and a 40 dB wideband amplifier located at the end of the optical fiber. The signal is then digitized using two channels of a Rohde+Schwarz RTO2044 oscilloscope. The 2.5 ns single photo-electron pulse width response of the PMTs corresponds to an input signal bandwidth of approximately 110 MHz. The oscilloscope was used in high definition mode, which provides an accuracy of 13 bits with an input bandwidth of 200~MHz.

Sampling frequencies between 250~Msamples/sec (250~MSps) and 1~GSps were considered for signal acquisition. The data was saved in records consisting of 10x 100~MS records, for a total recording time of 1sec @ 1~GSps to 4~sec @ 250~MSps. The 4~GB records were transferred via USB3 from the oscilloscope to an external solid state disk (SSD), resulting in a duty cycle of approximately 10\% at 500~MSps. Since the rate of data transfer to disk remains more or less constant independent of sampling rate, the duty cycle and thus the observation time in the sensitivity calculations scales with the sampling rate. The achievable bandwidth also depends on the sampling rate. As a result, the sensitivity is theoretically independent of the sampling rate chosen, since an increase in duty cycle is offset by a decrease in bandwidth. In practice, lower bandwidths (sampling rates) are less sensitive to delay differences due to staggering of mirror tiles, delay calculations, etc.

Before committing to observing on site, different sampling rates and the implementation of offline analysis software were investigated using an optical signal model in the laboratory. A test signal was generated using a Nichia laser diode which normally is used to produce coherent light at 405~nm. The diode was used as an LED at currents well below the laser threshold of the diode, but capable of fast modulation. A DC current was used to produce thermal emission, and an additional RF noise source with a 450~MHz high pass filter was used to add an additional optical component with a short random temporal correlation. This mimics the expected optical signal whose temporal coherence is also much shorter than the response time of the PMTs.

\begin{figure}
	\includegraphics[width=\columnwidth]{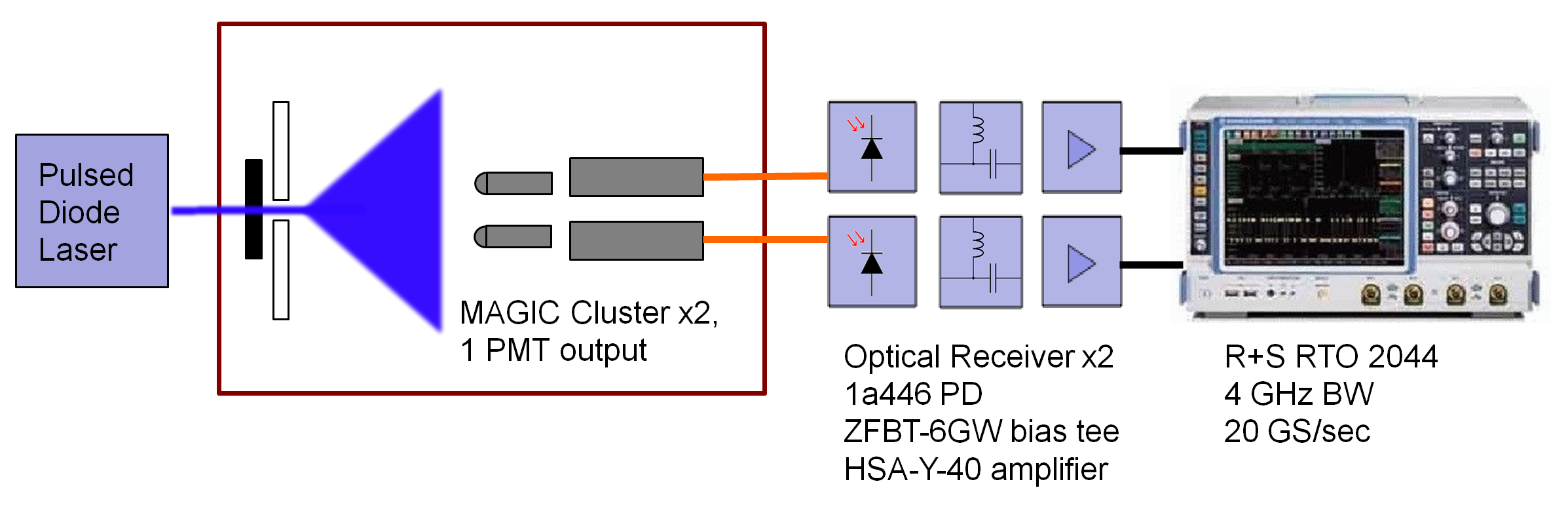}
    \caption{Sketch of the laboratory setup that was used to test the readout.}
    \label{fig:lab_setup}
\end{figure}
This test signal was used to verify the data analysis software and compare different sampling rates by means of the operation of two optical modules as used in the MAGIC cameras in a dark box with connection to the acquisition setup described (see figure \ref{fig:lab_setup}). 

A low pass flat time delay filter with a cutoff frequency of 117~MHz (Mini circuits SBLP-117+) was added to each channel between the amplifier and the oscilloscope to match the phase and frequency response of both channels and as a precaution against external interference signals.

\section{Analysis method}
\label{sec:analysis}

\subsection{Contrast and Pearson's correlation function}

As mentioned above the electronic chain of the MAGIC telescopes is AC coupled. Frequencies below 10 kHz are greatly attenuated. This means that
we do not know the DC value of $I_1$ or $I_2$. Therefore we can not calculate an absolute value for $g^{(2)}$ or $c$.

However our goal is not to measure $c$ but rather to determine the curve $c(d)/c(0)$ vs $d$ and fit it to Eq. \ref{eq:bessel} in order to extract the star's angular diameter. As we shall see we can extract this curve even if we do not know <I$_i$>.

To begin with we measure the Pearson's correlation coefficient $\rho$:
\begin{equation}
  \rho(\tau) = \frac{ < (I_1(t)-<I_1>) (I_2(t+\tau)-<I_2>) > } { \sqrt{<(I_1(t)-<I_1>)^2>} \sqrt{<(I_2(t)-<I_2>)^2>} }
  \label{eq:rho_definition}
\end{equation}
because I$_i$(t)-<I$_i$> are fast variations with respect to the pedestal i.e. they correspond to our actual measurement. We can use $\rho$ and its uncertainty to establish the strength of the correlation signal.

On the other hand $<I_1>$ and $<I_2>$ are respectively proportional to the anode current of the photodetectors (conventionally referred to as $DC_1$ in MAGIC-1 and $DC_2$ in MAGIC-2) with a fixed proportionality factor as long as the gains of the PMTs remain constant:

<I$_i$> $\propto$ DC$_i$

and the fluctuations of I$_1$ and I$_2$ are Poissonian, that is, proportional to the square root of the anode currents:

$\sqrt{<(I_i(t)-<I_i>)^2>}$ $\propto$ $\sqrt{DC_i}$

As a consequence:
\begin{equation} 
  c = K \frac{\rho}{\sqrt{DC_1 DC_2} }
  \label{eq:c_and_rho}
\end{equation}
where $K$ is constant if the gain of the PMTs (i.e. the high voltage) remain constant. During our observations the gain actually changed so we must write:
\begin{equation} 
  c = K \frac{\rho \sqrt{G_1 G_2}}{\sqrt{DC_1 DC_2} }
  \label{eq:c_and_rho_gains}
\end{equation}
where $G_1$ and $G_2$ are factors that are obtained from dedicated sets of DC measurements at different HV settings: see table \ref{tab:log}. 

DC$_1$ and DC$_2$ hardly change in time scales of a few seconds but depend strongly on the star and observation conditions on a longer time scale. In MAGIC DC$_1$ and DC$_2$ are measured roughly every second. We will calculate $\rho$ on the shortest possible time scales (subruns). To combine the data on longer time scales we will use the DC$_1$ and DC$_2$ measurements to determine $c/K$. The same procedure was used in the Narrabri observations (see chapter 10 of \cite{HB1974}).

The final goal is to determine $c(d)/K$ for a given star over a range of baselines. Using Eq. \ref{eq:cd_c0} would then allow us to determine the star's angular diameter.

\subsection{Computational method}

The computation time is driven by the calculation of the Pearson's coefficient $\rho$. 
Each data subrun is split into $N$ non-overlapping time windows of a fixed number of samples, $S$. Within each time window, the time is discretized, and runs from i=$1$ to $S$, therefore the signals are given by $I_1(i)$ and $I_2(i)$. For each one of these time windows, $\rho$ is estimated by applying Eq. \ref{eq:rho_definition} to its samples.  Subsequently, the final estimate of $\rho$ for a single data run is obtained by averaging the result obtained for each time window:

\begin{equation}
\rho(\tau)=\frac{1}{N} \sum_{i=1}^{N} \rho_i(\tau)
\label{eq:run_estimate}
\end{equation} 

where $\rho_i$ is the Pearson's coefficient computed for time window $i$. In addition,  the statistical error of $\rho(\tau)$ is estimated as:

\begin{equation}
\sigma(\rho(\tau)) = \frac{\Delta(\rho_i(\tau))}{\sqrt{N}}
\label{eq:error_run_estimate}
\end{equation}

where $\Delta(\rho_i(\tau))$ is the RMS of the sample of the $\rho$ estimates obtained from the $N$ time windows.

The only non-trivial averaging that has to be computed in the definition of $\rho$ (\ref{eq:rho_definition}) to obtain Eq. \ref{eq:run_estimate} is the cross term $<I_1(t) I_2(t+\tau)>$. For non-overlapping windows, it is obtained as

\begin{equation}
<I_1(t) I_2(t+\tau)> = \frac{\sum_{i=1}^{S} I_1(t_i) I_2(t_i+\tau) W(t_i+\tau)}{\sum_{i=1}^{S} W(t_i+\tau)}
\label{eq:time window}
\end{equation}

where $S$ is the number of samples; $\tau$ is given in discrete sample units; and $W(t)$ is a windowing function that is identical to one if and only if $t$ is within the time window and zero otherwise. The windowing function must be introduced to keep the computation of the estimate independent between different time windows. As a drawback, the statistics available for each value of $\tau$ is different, as it is reflected in the denominator of Eq. \ref{eq:time window}, which can be trivially computed to give

\begin{equation}
\sum_{i=1}^{S} W(t_i+\tau) = S - |\tau|
\end{equation}

where $|x|$ denotes the absolute value of $x$.

The direct computation of the numerator of \ref{eq:time window} for all non-trivial values of $\tau$ within a time window is time consuming, therefore we make use of the Convolution Theorem for discrete Fourier Transforms \cite{convolution}, that can be formulated as that given two periodic discrete functions $f$ and $g$, with period $P$, then 

\begin{equation}
\frac{1}{P}\sum_{i=1}^{P} f(i) g(i+j) = \mathcal{F}^{-1} (\mathcal{F} (f) \mathcal{F}^{*} (g))
\label{eq:fft}
\end{equation}

where $\mathcal{F}(x)$ denotes discrete Fourier's transform of $x$, $x^*$ denotes complex conjugate of $x$, and $\mathcal{F}^{-1}$ denotes the inverse discrete Fourier's transform. Moreover, the discrete Fourier's transforms can be efficiently computed using any of the Fast Fourier Transform (FFT) family of algorithms.

In order to make use of equation \ref{eq:fft}, it is necessary to adapt equation \ref{eq:time window}. For a given time window this is achieved by defining two periodic functions $\tilde{I}_1(i)$ and $\tilde{I}_2(i)$ of period $2S$, identically to $I_1(i)$ and $I_2(i)$ respectively for $i\  mod\ 2S \leq S$ and $0$ for $i\ mod\ 2S > S$. With this definition, if $\tau$ is restricted to $|\tau|<S-1$, the equation \ref{eq:time window} can be written as

\begin{equation}
<I_1(t) I_2(t+\tau)> = \frac{S}{S-|\tau|} \mathcal{F}^{-1} (\mathcal{F} (\tilde{I}_1) \mathcal{F}^{*} (\tilde{I}_2))
\end{equation}

which is the final procedure used to compute the cross term in the $\rho_i$ estimates in Eq. \ref{eq:run_estimate}. 

The size of the time windows used in the computation of $\rho$ obtained by optimizing the time required by the analysis. With the computational architecture and the FFT algorithm we have used, the optimal has been found to be $S=16384$ samples.
 
\subsection{Correction for the Moon}

The registered DCs are in fact produced by the combination of diffuse Moonlight and light coming from the star. 
During our actual observations (described in section \ref{sec:observations}) the DC produced by the Moon
are up to $\sim$15\% of the DC of the star. 

We will subtract the DCs produced by the Moon in each individual telescope from the measured DCs to calculate the DCs expected only from the star. We will apply the latter to equation \ref{eq:c_and_rho_gains} to determine the normalized contrast due to the star alone.

In fact, except for a single measurement on the first night we did not measure systematically the DCs produced only by the Moon. Instead we used a model to calculate the Moon DCs that has been validated with previous VHE Moon observations \citep{britzger}.

\section{Sensitivity estimate and selection of target stars}
\label{sec:selection}

An observational campaign was planned for the period around the Full Moon of April 2019, that is, during nights when the sky brightness is too high to schedule regular VHE observations.

\begin{table}
	\centering
	\caption{Setup parameters used to estimate the expected detection times:
	mirror area, QE of PMTs, QE of remaining optics, effective cross-correlation electrical bandwidth, normalized spectral distribution of the light and inverse of excess noise factor of the PMT. We assume that $b_v$ is similar to the electrical bandwidth of the individual channels.}
	\label{tab:parameters}
	\begin{tabular}{|cc|} 
		\hline
        $A$ & 250 m$^2$ \\
        \hline
        $\alpha$ & 29.5\% \\
        \hline
        $q$ & 0.236 \\
        \hline
        $b_v$ & 110 MHz \\
        \hline 
        $\sigma$ & 0.87 \\
        \hline  
        $F^{-1}$ & 0.87 \\
        \hline
    \end{tabular}
\end{table}
We can calculate the significance of the measured signal and the sensitivity of our setup using exclusively $\rho$. Equation 5.17 in \cite{HB1974} allows to calculate the significance (signal over noise) of the correlation for a given experimental setup and unpolarized light. The equation can be written as: 
\begin{equation}
S/N = \\
A \cdot \alpha(\lambda_0) \cdot q(\lambda_0) \cdot n(\lambda_0) \cdot |V|^2(\lambda_0,d) \cdot \sqrt{b_v} 
\cdot F^{-1} \cdot \sqrt{T/2} \cdot \sigma 
\label{eq:significance}
\end{equation}
where A is mirror area, $\alpha(\lambda_0)$ is the quantum efficiency (QE) of the PMTs for the filter's central wavelength $\lambda_0$, q$(\lambda_0)$ is the QE of the remaining optics, $n(\lambda_0)$ is the star's differential photon flux, $b_v$ is the effective cross-correlation electrical bandwidth, $F$ is the excess noise factor of the PMT and $T$ is the observation time. 
Finally $\sigma$ is the normalized spectral distribution of the light after the filter as defined in formula (5.6) of \cite{HB1974}. $\sigma$ would be equal to 1 if the filter transmission curve is a boxcar function and the spectrum of the light is flat whereas it reduces to 0.87 for the transmission curve in figure \ref{fig:effective_transmission}. 

We have assumed that both telescopes are identical and we have neglected the effect of the Moon and additional noise in the readout chain. As $\rho$ is proportional to $c$ for short time periods we may use the same expression to calculate the significance of $\rho$ for a subset of data of a few seconds.

We used equation \ref{eq:significance} to calculate the expected detection time of candidate stars. Table \ref{tab:parameters} shows the parameters of our setup. 

\begin{table*}
	\centering
	\caption{Brightness and angular diameter of the three best candidates for interferometric observations, together with the zenith angle and baselines of the observation campaign, the expected visibility, and the expected and measured time for a 5$\sigma$ detection signal.}
	\label{tab:estimate_significance}
	\begin{tabular}{|ccccccccc|} %
		\hline
        Star & m$_B$ & $\theta$ & Time & zd & $d$ & $V^2$ & Expected & Measured \\
             &     & (mas) & (UTC) & (deg) & (m) & & T$_{5\sigma}$ (sec) & T$_{5\sigma}$ (sec) \\
        \hline
        Adhara, $\epsilon$ CMa & 1.29 & 0.77$\pm$0.05 & 21:00 & 67 & 34.1 & 0.80 & 36 & 38~/~35\\
                                             & & & 22:00 & 76 & 25.9 & 0.88 & 54 & 57~/~64\\
                                             & & & 23:00 & 86 & 28.0 & 0.88 & 2160 & -\\
        \hline
        Benetnasch, $\eta$ UMa & 1.67 & 0.82$\pm$0.03 & 21:00 & 53 & 55.3 & 0.50 & 126 & - \\
                                                 & & & 22:00 & 42 & 66.0 & 0.36 & 210 & - \\
                                                 & & & 23:00 & 33 & 73.2 & 0.27 & 336 & 354 \\
                                                 & & & 00:00 & 25 & 78.6 & 0.22 & 510 & - \\
        \hline
Mirzam, $\beta$ CMa & 1.73 & 0.50$\pm$0.03 & 21:00 & 64 & 41.3 & 0.87 & 60 & 73 \\
 & & & 22:00 & 74 & 39.8 & 0.88 & 102 & 146 \\
 & & & 23:00 & 86 & 44.5 & 0.85 & 9780 & - \\
        \hline
    \end{tabular}
\end{table*}

We used the Jean-Marie Mariotti Center (JMMC) Stellar Diameter Catalog \citep{JSDC} to identify the brightest stars in B observable from La Palma during that period and having angular diameters between 0.3 and 1.0 mas so as to maximize the expected correlation signal. Table \ref{tab:estimate_significance} shows the time that is necessary to reach a signal of 5$\sigma$ significance following equation \ref{eq:significance} for the three best candidate stars. The differential flux after extinction $n$ was calculated from the magnitude in \cite{ducati} and the atmospheric extinction calculated for ORM and the zenith angle of the star. We have assumed an extinction coefficient k=0.23 mag/airmass for the filter's wavelength based on \cite{extinction}. $d$ has been calculated from the positions of the telescopes and the star's zenith and azimuth angles. 


The best candidates for a detection are Adhara ($\epsilon$ CMa), Benetnasch ($\eta$ UMa) and Mirzam ($\beta$ CMa). The next three stars are $\eta$~CMa, $\delta$~Sco and $\gamma$~Crv but the expected detection times exceeded 5 minutes. Since we were uncertain about a few of the setup parameters ($b_v$ or $F^{-1}$ for the specific central PMTs) we decided to take data only on Adhara, Benetnasch and Mirzam.

Unfortunately there were no stars of similar brightness with significantly smaller $\theta$. As we shall see this prevented us from making an absolute measurement of $\theta$ for the observed stars.

\section{Observations}
\label{sec:observations}

We installed the interferometry setup in MAGIC and took data on the three selected stars for five nights with bright and Full Moon in April 2019. A log of the observations can be found on table \ref{tab:log}. 

\begin{table*}
	\centering
	\caption{Log of the interferometric observations made with MAGIC in April 2019. Tabulated are the start day of the observations, the star name, time in UTC, general weather conditions, sampling rate in MSamples per second, anode current and High Voltage in the central pixels of MAGIC-1 (M1) and MAGIC-2 (M2).}
	\label{tab:log}
	\begin{tabular}{|ccccccccc|} 
		\hline
    Campaign &    Date   &	 	Star&	Time   &  Sampling  &	M1 DC  &	M2 DC & M1 HV  &	M2 HV \\
    day &         MJD  &	        	&	 (UTC) &    rate &	     ($\mu$A) &	   ($\mu$A) &  (V)      & (V)\\
		\hline
    1 &    2019/04/15 	 &	Adhara &	21:59 - 22:34 &	500 MSps 	&	37 - 21 & 33 - 15 &	1021 &	1063\\
      &   MJD 58588	 &	Benetnasch &	22:47 - 23:58 &	... 	&	42 - 44 &	41 - 44 & 961 &	983\\
		\hline
    2 & 2019/04/16 	 &	Adhara	& 21:11 - 21:41  &	... & unstable  & unstable & 1021 &	1063\\
      &   MJD 58589 &	Benetnasch & 21:45 - 23:53  &	... & unstable & unstable & 961 &	983\\
		\hline
    3 & 2019/04/17   	&	Adhara 	&21:02 - 22:51  & ... &	39 - 6 & 35 - 7 &	911 &	954\\
     & MJD 58590    &	Benetnasch &	22:55 - 23:58  & ... &	40 - 44 & 39 - 42 &	961 &	983\\
		\hline
    4 & 2019/04/18   	&	Adhara 	& 20:55 - 22:46  &	250 MSps &	43 - 7 & 43 - 8 &	931 &	983\\
      & MJD 58591        &	Benetnasch &	22:50 - 23:58  & ... & 	46 - 43 & 42 - 30 &	961 &	983\\	
		\hline
    5 & 2019/04/19  	 &	Mirzam &	20:56 - 22:38  &	... &	 	44 - 9 & 40 - 10 & 	1021 &	1063\\
	  & MJD 58592     &	Benetnasch	&22:43 - 23:59  &	... &	45 - 44 & 42 - 30 &	961 &	983\\
		\hline
	\end{tabular}
\end{table*}

The weather conditions were poor during the second night of observations so in what follows we will ignore the two data samples taken on that night.

Data were taken with two sampling rates: 250 MSps and 500 MSps to investigate the effect of the sampling rate in the sensitivity. We recorded a total of 8.2 TB of data over the whole campaign.

Different sets of PMT high voltages HV were applied with the goal to keep the DCs below a safe value of $\sim$40~$\mu$A but as high as possible to improve the PMT single photo-electron response and to reduce the impact of electronic noise.

In segmented acquisition mode the oscilloscope takes data of both channels simultaneously for 100 million samples and then stops digitizing while it saves the data to internal memory. Ten such ''subruns'' of 100 Msamples are successively recorded and these ten subruns are saved together as a ''run'' to the SSD adding even more dead time. The global dead time is about 90.7\% for 500 MSps sampling date and 81.4\% for 250 MSps. In other words, the duty cycle is correspondingly about 10\% and 20\%.

Adhara and Mirzam are observed long after transit so the DC strongly decreases along the observations whereas Benetnasch is observed during transit and the DC is roughly constant.

\section{Analysis results}
\label{sec:results}

\subsection{Significance of the correlation signals}

We have calculated Pearson's correlation coefficient $\rho$ for all data subruns as a function of time delay $\tau$.

\begin{figure}
	\includegraphics[width=\columnwidth]{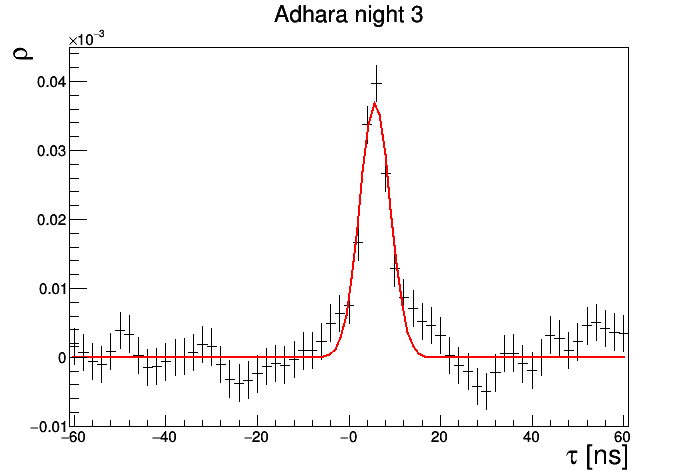}
    \caption{$\rho$ as a function of $\tau$ between the two telescopes for Adhara's observation on the third night. The variable delay term due to telescope orientations and positions has been subtracted. A signal with a significance of around 15$\sigma$ is observed near zero delay for an effective observation time of $\sim$600 seconds.
    }
    \label{fig:adhara_3_pearsons}
\end{figure}

Once the variable term of $\tau$ is corrected for, a clear correlation signal is detected in all samples at a fixed $\tau\sim$4~ns. As an illustration figure \ref{fig:adhara_3_pearsons} shows the mean $\rho$ for one of Adhara's data samples. The significance can be estimated either with respect to the RMS of $\rho$ far from the signal region or with respect to the RMS of $\rho$ calculated by dividing the total sample in smaller sub-samples. Both methods are consistent. For this specific data sample the significance is 15.4$\sigma$.

\begin{table}
	\centering
	\caption{Significance of the correlation signal measured for the four nights with good sky conditions.}
	\label{tab:significances}
	\begin{tabular}{|cccc|} 
		\hline
        Star & Campaign & Eff. Time & Significance  \\
         & day & (sec) & ($\sigma$) \\
        \hline  
        Adhara & 1 & 200 & 5.3 \\
        \hline 
        Adhara & 3 & 600 & 15.4\\
        \hline
        Adhara & 4 & 1040 & 12.6\\
        \hline
        Benetnasch & 1 & 402 & 5.8\\
        \hline
        Benetnasch & 3 & 350 & 4.9\\
        \hline
        Benetnasch & 4 & 760 & 7.3\\
        \hline
        Benetnasch & 5 & 880 & 5.1\\
        \hline
        Mirzam & 5 &  1072 & 9.3\\
		\hline
	\end{tabular}
\end{table}
Table \ref{tab:significances} tabulates the significance of the correlation signal measured for all data samples. 

Both Adhara and Mirzam were setting. The zenith angle increases quickly but the baseline hardly changes. As a consequence $|V|^2$ only changes by $\sim$10\% for Adhara and even less for Mirzam. In the contrary the baseline and consequently $|V|^2$ of Benetnasch change by almost a factor 2 during the first 2 hours of the night. We only targeted this star after 22:30 UTC because we prioritized Adhara and Mirzam for which we expected a stronger signal. In conclusion the observations cover only a small range in $|V|^2$ for each individual star and differences in detection time for each of them are dominated by changes in extinction, i.e. depend on zenith angle.

For Adhara and Mirzam the predicted times are in the order of 1 minute at the beginning of the night and degrade to tens of minutes in a matter of two hours. For Benetnasch the times are longer but the dependence with zenith angle is milder.

Let us compare the strength of the signal with the expectations in table \ref{tab:estimate_significance}. Since the detection times changes rather quickly along the night we make the comparison for short samples around the times in the table and we restrict ourselves to the data samples with more statistics, that is, Adhara nights 3 and 4, Benetnasch night 4 and Mirzam. The times that were needed to measure a 5$\sigma$ signal are tabulated in the last column of table \ref{tab:estimate_significance}. We have generally underestimated the detection times, but the expected and measured times match within 20\%. The only exception, a measured time 40\% longer than the expected one, is found for Mirzam around 22:00 UTC. Considering that we have not measured some of the parameters affecting the sensitivity (bandwidth, noise factor, atmospheric extinction, etc) the match is rather satisfactory.

Hanbury-Brown (see equation 11.12 and fig. 11.10 of \cite{HB1974}) concluded that the significance of an unresolved star at 45$^{\circ}$ elevation measured with the Narrabri interferometer scales with observation time and magnitude in $B$-band $m_B$ as: 
\begin{equation}
S/N = 0.4 \cdot \sqrt{T_0} \cdot 10^{-0.4\cdot m_B}
\label{eq:narrabri_sensitivity}
\end{equation}
where $T_0$ is the time of observation in seconds. A comparison with our results shows that our current interferometer is about 10~times more sensitive than Narrabri assuming that our duty cycle would be 100\%.

As a general indication, with this sensitivity we would detect a 5$\sigma$ signal from a $m_B$=5 mag unresolved star in about 3 hours. Resolving the star however demands to sample the correlation function down to, say, $|V|^2$=0.3. For a $m_B$=4 mag star this would be possible in about 5 hours.

It must be said that our longest observation was 15 minutes and we only targeted a limited number of stars so we have little control over our systematic errors and this sensitivity remains somewhat speculative.

\subsection{Contrast and correlation function}

We have used expression \ref{eq:c_and_rho} to calculate $c$ in arbitrary units for all runs. 

\begin{figure}
	\includegraphics[width=\columnwidth]{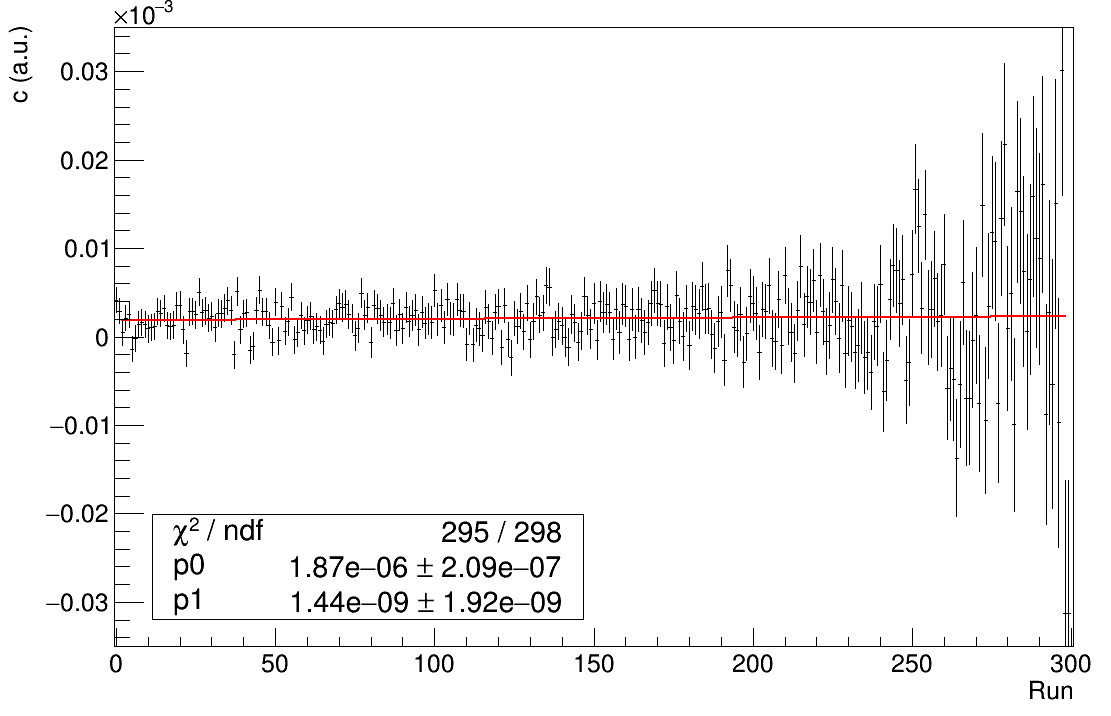}
    \caption{Contrast c as a function of run number for Adhara's observation on the third night. The units in the Y-axis are arbitrary. $p0$ and $p1$ are the free parameters in the linear fit $c$ = $p0$ + $p1 \cdot Run$.
    }
    \label{fig:adhara_3_c_run}
\end{figure}
Figure \ref{fig:adhara_3_c_run} shows $c$ (in arbitrary units) as a function of run number for Adhara and the third night. A fit to a straight line shows that there is no significant dependence on run number even if DC decreased by a factor $\sim$5 during the observations. This is a good crosscheck that equation \ref{eq:c_and_rho} can be used to estimate $c$. The decrease in DC and correspondingly photon flux from the star also results in an increase in the uncertainty of $c$.

A linear fit was applied to all samples and used to estimate $c$ for each star and night. All fits were successful although a small dependence with run number was still left in a few samples. This probably indicates that the effect of the Moon has not been fully removed.

\begin{figure}
	\includegraphics[width=\columnwidth]{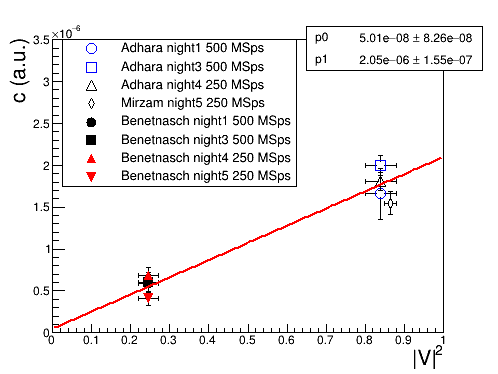}
    \caption{Contrast $c$ calculated from the data with equation \ref{eq:c_and_rho} as a function of $|V|^2$ determined from the baselines during the observations and star diameters in table \ref{tab:estimate_significance}. $p0$ and $p1$ are the free parameters in the linear fit $c$ = $p0$ + $p1 \cdot |V|^2$.}
    \label{fig:all_c_vs_gamma2}
\end{figure}
Even if we do not know $c$ in absolute terms we can crosscheck if it correlates linearly with $|V|^2$ for all stars and data samples under study, as expected from \ref{eq:cd_c0}. 

Figure \ref{fig:all_c_vs_gamma2} shows $c$ as a function of $|V|^2$ for all stars and observation nights. We have made a linear fit to the points. The results of the fit are shown on the same figure.

For each star $c$ is constant within the statistical error. This is consistent with the fact that the baseline hardly changes during the observations. 

The linear fit is successful and the fit goes through the origin, as expected from \ref{eq:cd_c0}. 

This plot may actually be used to calibrate our setup and measure the diameter of other stars. All in all these results are encouraging and indicate that MAGIC can already be used as an intensity interferometer. 

However it must be stressed that we have only taken a few minutes of data on a few nights of observations so we have hardly any control over our systematics. Only to mention a few effects we may suffer from sporadic electronic noise, telescope tracking may be unstable for some positions of the sky or the time delay between the two telescopes may slightly change over time scales of days. 

We must quantify these effects and in general take longer data samples on a number of stars up to at least magnitude 4, sampling the whole sky and different observation conditions, before producing scientific results.

\section{Conclusions and outlook}

We have searched for time correlation in the fluctuations of the signals produced by three stars of $\theta\sim$0.5-1 mas in the central PMTs of the two MAGIC telescopes. 

We have observed a clear correlation signal at different telescope baselines and four different observation nights. The significance of the signal is consistent with the parameters of our instrumental setup. 

We have also estimated the time correlation function for all data samples and compared with estimates based on the star diameter and telescope baselines. They are both consistent within the systematics of our instrument.

These results prove that MAGIC can be used as an optical intensity interferometer. Even with this simple instrumental setup we expect to detect a significant correlation signal for an unresolved star of m$_B \sim$5 in 3~hours of observations. 

Our setup is currently limited by the duty cycle of the acquisition system so our plan is to upgrade it to a system with a duty cycle in excess of 90\% using commercial digitizing cards that allow to transfer the data continuously to a computer. A higher duty cycle will generate a significant amount of data though. We aim at reducing the data online either at the digitizing cards or at the computer. 

We also plan to improve our mechanical setup. For safety reasons the current filter frame must be mounted and dismounted during the day so we are essentially limited to observing during Full Moon nights when VHE observations are impractical. We will build a new frame that can be deployed at any time. This allows not only to take additional data interspersed with regular VHE observations but also to test the response of the interferometer under very different light conditions.

On a longer time scale the interferometer can be improved in a number of ways. Since we digitize the signal using the standard photosensors already available in the VHE camera and the standard electronic path down to a central electronics room we can easily increase the number of channels. We are considering adding other spectral channels and two polarizations. 

PMTs can also be easily upgraded because the cameras of the MAGIC telescopes are modular. Each individual ''cluster'' of 7 PMTs is designed to be easily swapped with a new cluster for maintenance. Clusters equipped with new photosensors may be built as long as they are mechanically and electrically compatible: in fact clusters based on SiPMs have been already tested \citep{hahn,guberman,rando}. SiPMs show higher photodetection efficiency. New photosensors can also be selected to improve the time resolution or to match the spectral response of new spectral channels.

\section*{Acknowledgements}

This work would have been impossible without the support of our colleagues of the MAGIC collaboration.
We would also like to thank the Instituto de Astrof\'{\i}sica de Canarias for the excellent working conditions at the ORM. The financial support of the German BMBF and MPG, the Italian INFN and INAF, the Spanish MINECO partly using ERDF (FPA2017-82729-C6-6-R and C6-3-R) is gratefully acknowledged. This work was also supported by the Spanish Unidad de Excelencia ``Mar\'{\i}a de Maeztu'' (MDM-2014-0369) and by the 2018 Leonardo Grant for Researchers and Cultural Creators from BBVA Foundation.







\bsp	
\label{lastpage}
\end{document}